# A description of transverse momentum distributions in $p + p$ collisions at RHIC and LHC energies

Jia-Qi Hui，Zhi-Jin Jiang [a]

College of Science, University of Shanghai for Science and Technology, Shanghai 200093, China

**Abstract.** It has long been debated whether the hydrodynamics is suitable for the smaller colliding systems such as $p + p$ collisions. In this paper, by assuming the existence of longitudinal collective motion and long-range interactions in the hot and dense matter created in $p + p$ collisions, the relativistic hydrodynamics incorporating with the nonextensive statistics is used to analyze the transverse momentum distributions of the particles. The investigations of present paper show that the hybrid model can give a good description of the currently available experimental data obtained in $p + p$ collisions at RHIC and LHC energies, except for $p$ and $\bar{p}$ produced in the range of $p_T > 3.0$ GeV/c at $\sqrt{s}$ =200 GeV.

## 1 Introduction

In the past decade, the experimental results of heavy ion collisions at both RHIC and LHC energies have been extensively studied. These studies have shown that the strongly coupled quark-gluon plasma (sQGP) might be created in these collisions [1-9], which exhibits a clear collective behavior almost like a perfect fluid with very low viscosity [10-28]. Therefore, the evolution of sQGP can be described in the scope of relativistic hydrodynamics. However, unlike heavy ion collisions, $p + p$ collisions are a relatively smaller system with lower multiplicity, larger viscosity, and larger fluctuation [29]. The reasonableness of applying relativistic hydrodynamics in depicting the evolution of sQGP created in $p + p$ collisions has been undergone an endless debate.

In this paper, by supposing the existence of collective flow in $p + p$ colliding direction, the relativistic hydrodynamics including phase transition is introduced to describe the longitudinal expansion of sQGP. Besides the collective flow, the thermal motion also exists in sQGP. The evolution of sQGP is therefore the superposition of collective flow and thermal motion. Known from the investigations of Ref. [30, 31], the long-range interactions might appear in sQGP. This guarantees the reasonableness of nonextensive statistics in describing the thermodynamic aspects of sQGP. Hence, in this paper, we will use the nonextensive statistics instead of conventional statistics to characterize the thermal motion of the matter created in $p+p$ collisions.

---

[a] Corresponding author. Email: Jzj265@163.com

The nonextensive statistics, *i.e.* Tsallis nonextensive thermostatistics, is the generalization of conventional Boltzmann-Gibbs statistics, which is proposed by C. Tsallis in his pioneer work of Ref. [32]. This statistical theory overcomes the inabilities of the conventional statistical mechanics by assuming the existence of long-range interactions, long-range microscopic memory, or fractal space-time constrains in the thermodynamic system. It has a wide range of applications in cosmology [33], phase shift analyses for the pion-nucleus scattering [34], dynamical linear response theory, and variational methods [35]. It has achieved a great success in talking many physical problems, such as the solar neutrino problems [36], many-body problems, and the problems in astrophysical self-gravitating systems [37].

The article is organized as follows. In section 2, a brief description is given to the employed hydrodynamics, presenting its analytical solutions. The solutions are then used in section 3 to formulate the transverse momentum distributions of the particles produced in $p+p$ collisions in the light of Cooper-Frye prescription. The last section 4 is about conclusions.

## 2 A brief introduction to the hydrodynamic model

The main content of the relativistic hydrodynamic model [15] used in this paper is as follows.

The expansion of fluid obeys the continuity equation

$$\partial_\mu T^{\mu\nu} = 0, \mu, \nu = 0, 1, \qquad (1)$$

where

$$T^{\mu\nu} = (\varepsilon + p)u^\mu u^\nu - pg^{\mu\nu} \qquad (2)$$

is the energy-momentum tensor of fluid, $g^{\mu\nu} = \text{diag}(1, -1)$ is the metric tensor. The four-velocity of fluid $u^\mu = (u^0, u^1) = (\cosh y_F, \sinh y_F)$, where $y_F$ is the rapidity of fluid. $\varepsilon$ and $p$ in Eq. (2) are the energy density and pressure of fluid, respectively, which are related by the sound speed $c_s$ of fluid *via* the equation of state

$$\frac{\mathrm{d}p}{\mathrm{d}\varepsilon} = \frac{s\mathrm{d}T}{T\mathrm{d}s} = c_s^2, \tag{3}$$

where, $T$ and $s$ are the temperature and entropy density of fluid, respectively.

The projection of Eq. (1) to the direction of $u_\mu$ leads to the continuity equation for entropy conservation

$$\partial_\nu(su^\nu) = 0. \tag{4}$$

The projection of Eq. (1) to the direction perpendicular to $u_\mu$ gives equation

$$\frac{\partial(T \sinh y_F)}{\partial t} + \frac{\partial(T \cosh y_F)}{\partial z} = 0, \tag{5}$$

which means the existence of a scalar function $\phi$ satisfying

$$\frac{\partial \phi}{\partial t} = T \cosh y_F, \frac{\partial \phi}{\partial z} = -T \sinh y_F. \tag{6}$$

By using $\phi$ and Legendre transformation, Khalatnikov potential $\chi$ can be introduced *via* relation

$$\chi = \phi - tT \cosh y_F + zT \sinh y_F, \tag{7}$$

which changes the coordinate base of $(t, z)$ to that of $(\omega, y_F)$

$$t = \frac{e^\omega}{T_0}\left(\frac{\partial \chi}{\partial \omega} \cosh y_F + \frac{\partial \chi}{\partial y_F} \sinh y_F\right), \tag{8}$$

$$z = \frac{e^\omega}{T_0}\left(\frac{\partial \chi}{\partial \omega} \sinh y_F + \frac{\partial \chi}{\partial y_F} \cosh y_F\right), \tag{9}$$

where $T_0$ is the initial temperature of sQGP, and $\omega = -\ln(T/T_0)$. In terms of $\chi$, Eq. (4) can be rewritten as the so-called equation of telegraphy

$$\frac{\partial^2 \chi}{\partial \omega^2} - 2\beta \frac{\partial \chi}{\partial \omega} - \frac{1}{c_s^2}\frac{\partial^2 \chi}{\partial y_F^2} = 0, \beta = \frac{1-c_s^2}{2c_s^2}. \tag{10}$$

With the expansion of created matter, its temperature becomes lower and lower. When the temperature drops from the initial temperature $T_0$ to the critical temperature $T_c$, phase transition occurs. This will modify the value of sound speed of fluid. In sQGP, $c_s = c_0 = 1/\sqrt{3}$, which is the sound speed of a massless perfect fluid, being the maximum of $c_s$. In the hadronic state, $0 < c_s = c_h \leq c_0$. At the point of phase transition, $c_s$ is discontinuous.

The solutions of Eq. (10) for sQGP and hadronic state are respectively [15],

$$\chi_0(\omega, y_F) = \frac{Q_0 c_0}{2} e^{\beta_0 \omega} I_0\left(\beta_0 \sqrt{\omega^2 - c_0^2 y_F^2}\right), \tag{11}$$

$$\chi_h(\omega, y_F) = \frac{Q_0 c_0}{2} S(\omega) I_0[\lambda(\omega, y_F)], \tag{12}$$

where, $I_0$ is the 0th order modified Bessel function, and

$$\beta_0 = (1 - c_0^2) / 2c_0^2 = 1, \quad S(\omega) = e^{\beta_h(\omega - \omega_c) + \beta_0 \omega_c}, \quad \lambda(\omega, y_F) = \beta_h c_h \sqrt{y_h^2(\omega) - y_F^2},$$

$$\beta_h = (1 - c_h^2) / 2c_h^2, \quad \omega_c = -\ln(T_c / T_0), \quad y_h(\omega) = [(\omega - \omega_c) / c_h] + (\omega_c / c_0). \tag{13}$$

The $Q_0$ in Eqs. (11) and (12) is a free parameter determined by fitting the theoretical results with experimental data.

## 3 The transverse momentum distributions of the particles produced in $p + p$ collisions

### (1) The energy of quantum of produced matter

The nonextensive statistics is based on the following two postulations [32, 36]

(a) The entropy of a statistical system possesses the form of

$$s_q = \frac{1}{q-1}(1 - \sum_{i=1}^{\Omega} p_i^q), \tag{14}$$

where $p_i$ is the probability of a given microstate among $\Omega$ ones, $q$ is a fixed real parameter. The defined entropy has the usual properties of positivity, equiprobability, irreversibility, and in the limit of $q \to 1$, it reduces to the conventional Boltzmann-Gibbs entropy

$$s = -\sum_i p_i \ln p_i. \tag{15}$$

(b) The mean value of an observable $\mathcal{O}$ is defined as

$$\bar{\mathcal{O}}_q = \sum_{i=1}^{\Omega} p_i^q \mathcal{O}_i, \tag{16}$$

where $\mathcal{O}_i$ is the value of an observable $\mathcal{O}$ in the microstate $i$.

From the above two postulations, the average occupational number of quantum in the state with temperature $T$ can be written in a simple analytical form [38]

$$\bar{n}_q = \frac{1}{[1 + (q-1)(E - \mu_B)/T]^{1/(q-1) + \delta}}. \tag{17}$$

Here, as usual, $E$ is the energy of quantum, and $\mu_B$ is its baryochemical potential. For baryons $\delta = 1$ and for mesons $\delta = -1$. In the limit of $q \to 1$, it reduces to the conventional Fermi-Dirac or Bose-Einstein distributions. Hence, the value of $q$ in the nonextensive statistics represents the degree of deviation from the conventional statistics. Known from Eq. (17), the average energy of quantum in the state with temperature $T$ reads

$$\bar{E}_q = \frac{m_T \cosh(y - y_F)}{\{1 + [(q-1)(m_T \cosh(y - y_F) - \mu_B)]/T\}^{1/(q-1) + \delta}}, \tag{18}$$

where, $y$ is the rapidity of quantum, $m_T = \sqrt{p_T^2 + m^2}$ is its transverse mass with rest mass $m$ and transverse momentum $p_T$.

**(2) The transverse momentum distributions of the particles produced in $p + p$ collisions**

With the expansion of hadronic matter, its temperature becomes even lower. As the temperature drops to the so-called kinetic freeze-out temperature $T_f$, the inelastic collisions among hadronic matter stop. The yields of produced particles keep unchanged becoming the measured results. According to Cooper-Frye scheme [39], the invariant multiplicity distributions of produced particles take the form as [15, 39]

$$\frac{d^2N}{2\pi p_T \, dy \, dp_T} = \frac{A}{(2\pi)^3} \int_{-y_h(\omega_f)}^{y_h(\omega_f)} (\cosh y \frac{dz}{dy_F} - \sinh y \frac{dt}{dy_F}) \bar{E}_q \Big|_{T=T_f} dy_F, \quad (19)$$

where $A$ is the area of overlap region of collisions, $\omega_f = -\ln(T_f / T_0)$, and the integrand takes values at the moment of $T = T_f$. The meaning of Eq. (19) is evident. The part of integrand in the round brackets is proportional to the rapidity density of fluid [39]. Hence, Eq. (19) is the convolution of rapidity of fluid with the energy of the particles in the state with temperature $T$. From Eq. (8) and (9)

$$\cosh y \frac{dz}{dy_F} - \sinh y \frac{dt}{dy_F}$$
$$= \frac{1}{T} c_s^2 \frac{\partial}{\partial \omega}\left(\chi + \frac{\partial \chi}{\partial \omega}\right) \cosh(y - y_F) - \frac{1}{T} \frac{\partial}{\partial y_F}\left(\chi + \frac{\partial \chi}{\partial \omega}\right) \sinh(y - y_F). \quad (20)$$

Substituting $\chi$ in Eq. (20) by the $\chi_h$ of Eq. (12) and taking the values at the moment of $T = T_f$, it becomes

$$\left(\cosh y \frac{dz}{dy_F} - \sinh y \frac{dt}{dy_F}\right)\Big|_{T=T_f}$$
$$= \frac{Q_0 c_0}{T_f} (\beta_h c_h)^2 S(\omega_f) \left[B(\omega_f, y_F) \sinh(y - y_F) + C(\omega_f, y_F) \cosh(y - y_F)\right], \quad (21)$$

where

$$B(\omega_f, y_F) = \frac{\beta_h y_F}{\lambda(\omega_f, y_F)} \left\{ \frac{\beta_h c_h y_h(\omega_f)}{\lambda(\omega_f, y_F)} I_0[\lambda(\omega_f, y_F)] + \left[\frac{\beta_h + 1}{\beta_h} - \frac{2\beta_h c_h y_h(\omega_f)}{\lambda^2(\omega_f, y_F)}\right] I_1[\lambda(\omega_f, y_F)] \right\}, \quad (22)$$

$$C(\omega_f, y_F) = \left\{ \frac{\beta_h + 1}{\beta_h} + \frac{[\beta_h c_h y_h(\omega_f)]^2}{\lambda^2(\omega_f, y_F)} \right\} I_0[\lambda(\omega_f, y_F)]$$
$$+ \frac{1}{\lambda(\omega_f, y_F)} \left\{ \frac{y_h(\omega_f)}{c_h} + 1 - \frac{2[\beta_h c_h y_h(\omega_f)]^2}{\lambda^2(\omega_f, y_F)} \right\} I_1[\lambda(\omega_f, y_F)], \quad (23)$$

where $\lambda(\omega_f, y_F) = \beta_h c_h \sqrt{y_h^2(\omega_f) - y_F^2}$, $I_1$ is the 1st order modified Bessel function.

By using Eqs. (19) and (21)-(23), we can obtain the transverse momentum distributions of produced particles as shown in Figs. 1, 2, and 3.

Fig. 1 shows the transverse momentum spectra of $K_S^0$, $K^+$, $K^-$, $\Lambda$, $\bar{\Lambda}$, $\Xi^-$, $\bar{\Xi}^+$, and $\Omega^-+\bar{\Omega}^+$ produced in $p+p$ collisions at $\sqrt{s} = 200$ GeV. The solid dots, circles, and solid triangles represent the experimental data of the STAR Collaboration [40]. The solid curves are the results calculated from Eq. (19). The values of free parameters $q$, $Q_0$, and $\chi^2$/NDF are listed in Table 1.

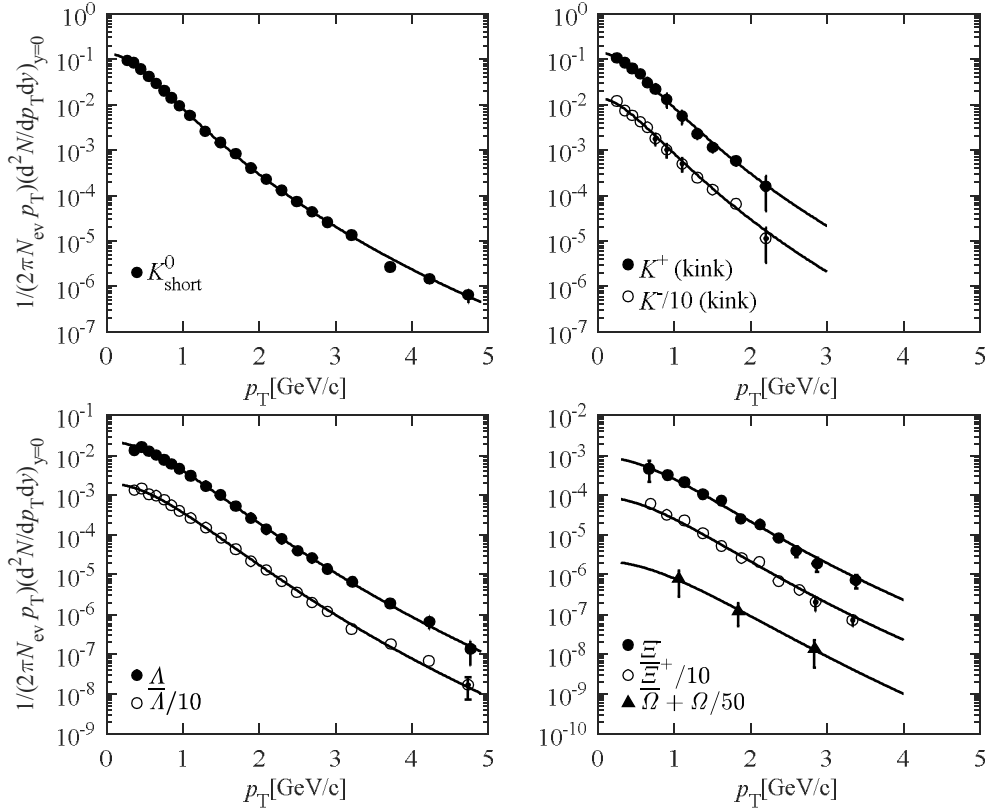

**Fig. 1.** The transverse momentum distributions of strange particles ($|y| < 0.5$) produced in $p+p$ collisions at $\sqrt{s} = 200$ GeV. The solid dots, circles, and solid triangles represent the experimental data of the STAR Collaboration [40]. The solid curves are the results calculated from Eq. (19).

Table 1. The values of $q$, $Q_0$, and $\chi^2$/NDF obtained from the analyses of STAR data [40] in $p+p$ collisions at $\sqrt{s} = 200$ GeV

| Parameters | $K_S^0$ | $K^+/K^-$ | $\Lambda/\bar{\Lambda}$ | $\Xi^-/\bar{\Xi}^+$ | $\Omega^-+\bar{\Omega}^+$ |
|---|---|---|---|---|---|
| $q$ | 1.083±0.002 | 1.083±0.005 | 1.062±0.001 | 1.075±0.003 | 1.068±0.007 |
|  |  | 1.083±0.006 | 1.062±0.001 | 1.075±0.003 |  |
| $Q_0$ | 0.084±0.005 | 0.087±0.012 | 0.379±0.026 | 0.026±0.006 | 0.016±0.010 |
|  |  | 0.086±0.012 | 0.337±0.023 | 0.026±0.005 |  |
| $\chi^2$/NDF | 0.68 | 0.32/0.39 | 0.47/0.90 | 0.47/0.64 | 0.02 |

Fig. 2 presents the transverse momentum spectra of $\pi^+, \pi^-, K^+, K^-, p,$ and $\bar{p}$ produced in $p + p$ collisions at $\sqrt{s} = 200$ GeV. The solid dots, solid triangles, solid squares, circles, triangles, and squares represent the experimental data of the PHENIX Collaboration [41]. The solid curves are the results calculated from Eq. (19). The values of free parameters $q$, $Q_0$, and $\chi^2$/NDF are summarized in Table 2. The theoretical model can give a good description of the experimental data for $\pi^+, \pi^-, K^+, K^-$ in the whole measured transverse momentum range, and for $p$ and $\bar{p}$ in the range of $p_T \leq 3.0$ GeV/c. In the range of $p_T > 3.0$ GeV/c, the deviation appears as shown in Fig. 3, which shows the transverse momentum distributions of $p$ and $\bar{p}$ in the whole measured $p_T$ range.

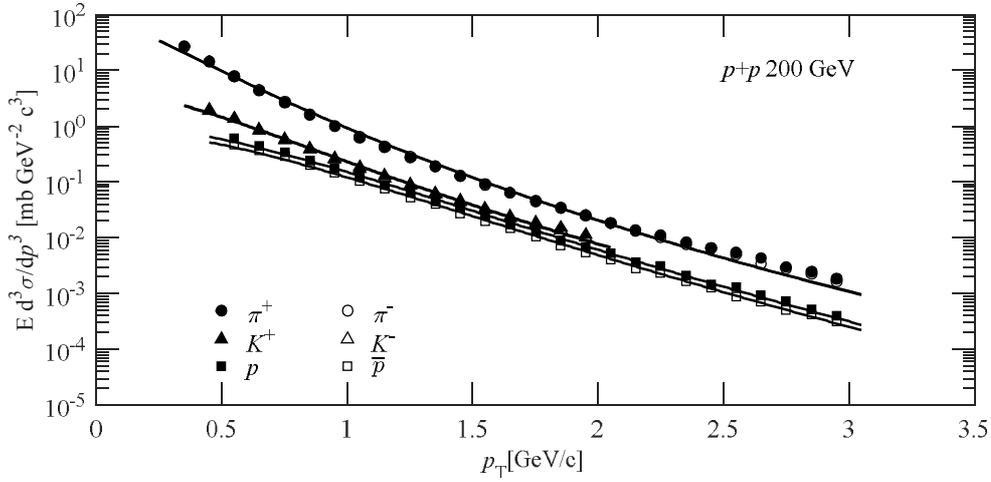

**Fig. 2.** The transverse momentum distributions of $\pi^+, \pi^-, K^+, K^-, p,$ and $\bar{p}$ produced in $p + p$ collisions at $\sqrt{s} = 200$ GeV at midrapidity. The solid dots, solid triangles, solid squares, circles, triangles, and squares represent the experimental data of the PHENIX Collaboration [41]. The solid curves are the results calculated from Eq. (19).

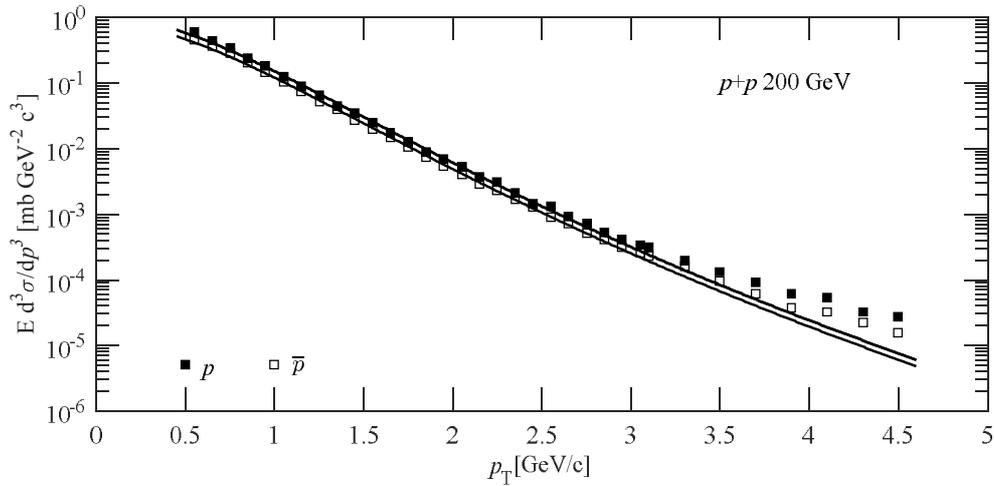

**Fig. 3.** The transverse momentum distributions of $p$ and $\bar{p}$ produced in $p + p$ collisions at $\sqrt{s} = 200$ GeV in the whole measured $p_T$ range. The solid squares and squares represent the experimental data of the PHENIX Collaboration [41]. The solid curves are the results calculated from Eq. (19).

Table 2. The values of $q$, $Q_0$, and $\chi^2$/NDF obtained from the analyses of PHENIX data [41] in $p+p$ collisions at $\sqrt{s} = 200$ GeV

| Parameters | $\pi^+/\pi^-$ | $K^+/K^-$ | $p/\bar{p}$ |
|---|---|---|---|
| $q$ | 1.075±0.003 | 1.080±0.003 | 1.060±0.002 |
|  | 1.075±0.003 | 1.080±0.003 | 1.060±0.001 |
| $Q_0$ | 10.439±0.015 | 3.699±0.005 | 13.099±1.551 |
|  | 10.342±0.015 | 3.602±0.006 | 10.479±1.221 |
| $\chi^2$/NDF | 5.25/3.28 | 1.57/1.01 | 0.47/0.17 |

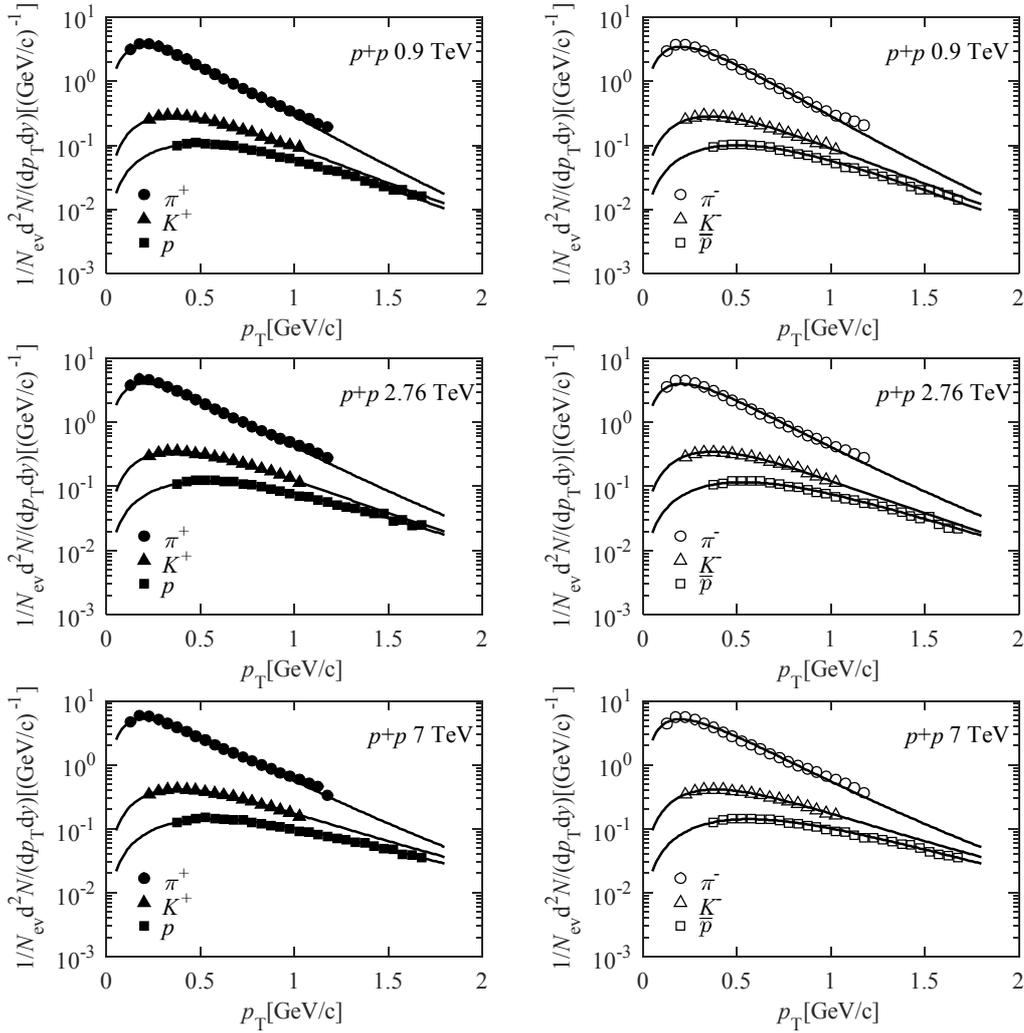

**Fig. 4.** The transverse momentum distributions of the identified charged particles ($|y| < 1$) produced in $p+p$ collisions at $\sqrt{s} = 0.9$, 2.76, and 7 TeV (from top to bottom). The solid dots, solid triangles, solid squares, circles, triangles, and squares represent the experimental data of the CMS Collaboration [42]. The solid curves are the results calculated from Eq. (19).

Fig. 4 shows the transverse momentum spectra of $\pi^+$, $\pi^-$, $K^+$, $K^-$, $p$, and $\bar{p}$ produce in $p+p$ collisions at $\sqrt{s} = 0.9$, 2.76, and 7 TeV. The solid dots, solid triangles, solid squares, circles, triangles, and squares represent the experimental data of the CMS Collaboration [42]. The solid curves are the results calculated from Eq. (19). The values of free parameters $q$, $Q_0$, and $\chi^2$/NDF are summarized in Table 3.

Table 3. The values of $q$, $Q_0$, and $\chi^2$/NDF obtained from the analyses of CMS data [42] in $p + p$ collisions at LHC energies

| $\sqrt{s}$ | Parameters | $\pi^+/\pi^-$ | $K^+/K^-$ | $p/\bar{p}$ |
|---|---|---|---|---|
| 0.9 TeV | $q$ | 1.064±0.002 | 1.090±0.002 | 1.071±0.001 |
| | | 1.064±0.003 | 1.090±0.002 | 1.071±0.001 |
| | $Q_0$ | 0.161±0.004 | 0.045±0.001 | 0.152±0.005 |
| | | 0.159±0.004 | 0.044±0.002 | 0.145±0.003 |
| | $\chi^2$/NDF | 8.50/10.85 | 0.37/0.47 | 1.10/1.84 |
| 2.76 TeV | $q$ | 1.078±0.002 | 1.100±0.002 | 1.088±0.001 |
| | | 1.078±0.002 | 1.100±0.002 | 1.088±0.001 |
| | $Q_0$ | 0.051±0.001 | 0.015±0.0004 | 0.037±0.001 |
| | | 0.050±0.001 | 0.014±0.0004 | 0.036±0.001 |
| | $\chi^2$/NDF | 8.90/8.25 | 0.30/0.62 | 1.75/1.35 |
| 7 TeV | $q$ | 1.084±0.002 | 1.120±0.003 | 1.105±0.001 |
| | | 1.084±0.003 | 1.120±0.002 | 1.105±0.001 |
| | $Q_0$ | 0.004±0.00004 | 0.001±0.00002 | 0.002±0.00004 |
| | | 0.004±0.00008 | 0.001±0.00002 | 0.002±0.00002 |
| | $\chi^2$/NDF | 7.28/7.69 | 0.23/0.35 | 1.13/1.43 |

In calculations, the sound speed in hadronic state takes the value of $c_h = 0.35$ [43, 44]. The critical temperature takes the value of $T_c = 0.16$ GeV [45]. For $\sqrt{s} = 200$ GeV, the initial temperature takes the value of $T_0 = 0.35$ GeV [46], the kinetic freeze-out temperature takes the values of $T_f = 0.12$ GeV for strange particles and pions, and for protons, $T_f = 0.13$ GeV from the investigation of Ref. [47], which also shows that the baryochemical potential takes the value of $\mu_B = 0.01$ GeV. For $\sqrt{s} = 0.9$, 2.76, and 7 TeV, referring to Ref. [46], the initial temperatures are estimated to be $T_0 = 0.4$, 0.6, and 1.5 GeV, respectively. The kinetic freeze-out temperature takes the values of $T_f = 0.12$ GeV for pions and kaons, and for protons, $T_f = 0.13$ GeV. The baryochemical potential takes the value of $\mu_B = 0$ [47].

The parameters $Q_0$ and $T_0$ have the same effects, they all affect the amplitudes of the theoretical curves. They are different from the parameter $q$ which affects the slopes of the theoretical curves.

## 4 Conclusions

By assuming the existence of longitudinal collective motion and long-range interactions in sQGP produced in $p+p$ collisions, the relativistic hydrodynamics including phase transition together with the nonextensive statistics is used to discuss the transverse momentum distributions of the particles produced in $p+p$ collisions at $\sqrt{s}=0.2, 0.9, 2.76$, and 7 TeV.

The theoretical model used in this paper contains a rich information about the transport coefficients of fluid, such as the sound speed $c_0$ in sQGP, the sound speed $c_h$ in hadronic state, the initial temperature $T_0$, the critical temperature $T_c$, the kinetic freeze-out temperature $T_f$, and the baryochemical potential $\mu_B$. Except for $T_0$, the other five parameters take the values either from the widely accepted theoretical results or from experimental measurements. As for $T_0$, there are no acknowledged values so far. In this paper, $T_0$ takes the values from other studies. The investigations of present paper show the conclusions as follows:

a) The theoretical model can give a good description of the currently available experimental data collected in $p+p$ collisions at RHIC and LHC energies with the only exception of $p$ and $\bar{p}$ measured in the range of $p_T > 3.0$ GeV/c at $\sqrt{s}=200$ GeV, which might be caused by the hard scattering process [48]. To improve the fitting conditions, the results of perturbative QCD should be taken into account.

b) The fitted values of $q$ are close to 1. This means that the deviation between nonextensive statistics and conventional statistics is small. While, it is this small difference that plays an essential role in fitting the experimental data.

## Conflict of Interests

The authors declare that there is no conflict of interests regarding the publication of this paper.

## Acknowledgments

This work is supported by the Shanghai Key Lab of Modern Optical System.## References

1. BRAHMS Collaboration (I. Arsene *et al.*), Nucl. Phys. A **757**, 1 (2005).

2. PHOBOS Collaboration (B. Back *et al.*), Nucl. Phys. A **757**, 28 (2005).

3. STAR Collaboration (J. Adams *et al.*), Nucl. Phys. A **757**, 102 (2005).10